\documentclass[aip,reprint]{revtex4-1}
\usepackage{graphicx}
\usepackage{ulem}
\usepackage{amssymb}
\usepackage{amsmath}
\usepackage{wasysym}
\usepackage{relsize}
\begin{document}


\newcommand{\be}{\begin{equation}}
\newcommand{\ee}{\end{equation}}
\newcommand{\bea}{\begin{eqnarray}}
\newcommand{\eea}{\end{eqnarray}}
\newcommand{\Tbar}{{\bar{T}}}
\newcommand{\En}{{\cal E}}
\newcommand{\K}{{\cal K}}
\newcommand{\U}{{\cal U}}
\newcommand{\GC}{{\cal \tt G}}
\newcommand{\Lop}{{\cal L}}
\newcommand{\DB}[1]{\marginpar{\footnotesize DB: #1}}
\newcommand{\q}{\vec{q}}
\newcommand{\kt}{\tilde{k}}
\newcommand{\Lopn}{\tilde{\Lop}}
\newcommand{\noi}{\noindent}
\newcommand{\ovn}{\bar{n}}
\newcommand{\ovx}{\bar{x}}
\newcommand{\ovE}{\bar{E}}
\newcommand{\ovV}{\bar{V}}
\newcommand{\ovU}{\bar{U}}
\newcommand{\ovJ}{\bar{J}}
\newcommand{\calE}{{\cal E}}
\newcommand{\ovphi}{\bar{\phi}}
\newcommand{\zt}{\tilde{z}}
\newcommand{\ttl}{\tilde{\theta}}
\newcommand{\nuv}{\rm v}
\newcommand{\ds}{\Delta s}
\newcommand{\fn}{{\small {\rm  FN}}}
\newcommand{\cc}{{\cal C}}
\newcommand{\cd}{{\cal D}}
\newcommand{\tth}{\tilde{\theta}}
\newcommand{\cb}{{\cal B}}
\newcommand{\cg}{{\cal G}}
\newcommand\norm[1]{\left\lVert#1\right\rVert}
\title{The anode proximity effect for generic smooth field emitters}
 

\author{Debabrata Biswas}
\affiliation{
Bhabha Atomic Research Centre,
Mumbai 400 085, INDIA}
\affiliation{Homi Bhabha National Institute, Mumbai 400 094, INDIA}


\begin{abstract}
  The proximity of the anode to a curved field electron emitter alters the electric field at the apex and its
  neighbourhood. A formula for the apex field enhancement factor, $\gamma_a(D)$,
  for generic smooth emitters is derived using the
  line charge model when the anode is at a distance $D$ from the cathode plane. The resulting approximately
  modular form is such that the  anode proximity contribution can be calculated separately
  (using geometric quantities such as the anode-cathode distance $D$,
  the emitter height $h$ and the emitter apex radius of curvature $R_a$)
  and plugged into the expression for $\gamma_a(\infty)$.
  It is also shown that the variation of the enhancement factor on the surface of the emitter close to the apex
  is unaffected by the presence of the anode and continues to obey the generalized cosine law.
  These results are verified numerically for various generic emitter shapes using
  COMSOL Multiphysics\textsuperscript{\textregistered}.
  Finally, the theory is applied to explain experimental observations on the scaling behavior of the $I-V$ field emission curve.
\end{abstract}

\maketitle

\section{Introduction}
\label{sec:intro}

Analytical models of curved field emitters aligned along the asymptotic electric field generally assume the
anode to be at infinity. The floating sphere at emitter plane potential, the line charge model and the point
charge model are examples where the anode is neglected as a first approximation.
In contrast, numerical modeling takes into explicit account the presence of
the anode with studies showing an increase in electrostatic field when the anode is in close proximity to the
emitter apex \cite{ev2002,forbes2003,wang2004,smith2005,podenok,pogo2009,pogo2010,filip2011,lenk2018}.

The phenomenon under investigation in all of the above is field enhancement at curved emitter tips.
The curvature leads to
a magnified electric field at the emitter apex, thereby lowering and narrowing  the potential barrier as
seen by a tunneling electron. Thus, field emission occurs from such emitter tips even at moderate macroscopic
fields of about $1$MV/m if the field is enhanced a few thousand times. The degree of enhancement
is measured in terms of the field enhancement factor $\gamma$ such that the local field on the
emitter is expressed as $E_l = \gamma E_0$ where $E_0$ is the macroscopic or asymptotic field far away from the
emitter. The presence of an anode in close
proximity to the emitter tip adds to the enhancement. Situations where the anode-emitter distance is small
are of relevance  in microscopy and lithography.

Of the models mentioned above, the floating sphere at emitter plane potential
is a well researched  simplified analytical model for carbon nanotubes. It over-predicts
the apex field enhancement factor when the anode is far away (denoted here by $\gamma_a(\infty))$ but the
method has been suitably adapted 
to deal with the anode plate at a finite distance $D$ ($> h$) from
the cathode using an infinite summation over image charge contributions from successive planes. It predicts
the apex field enhancement factor (AFEF) for an anode-cathode separation $D$ to be \cite{wang2004,pogo2010}

\be
\gamma_a(D) = \gamma_a(\infty) + \zeta(3) \big(\frac{h}{D}\big)^3
\ee

\noi
provided $h/D \lessapprox 0.7$. Here  $\gamma_a(\infty) \simeq  h/R_a + 7/2$
where $h$ is the emitter height and $R_a$ the apex radius). Notably, the correction factor due to anode proximity is independent
of the apex radius of curvature as per the predictions of this model.

The line charge model  \cite{pogo2009,harris15,jap16} replaces the curved axially symmetric emitter by a line charge which can
be thought of as the projection of surface charges that accumulate  on the surface of an actual emitter
due to the termination of field lines. The line charge together with the macroscopic electrostatic field $E_0$ 
gives rise to a zero potential surface which corresponds to the actual curved emitter surface. The model is
in principle applicable to all curved emitters, each having a unique line charge density that is
in general nonlinear.

The model accurately describes a curved emitter mounted on a cathode plane
in a diode configuration. Recent studies \cite{db_fef} for a general smooth line charge density $\Lambda(s)$
show that the apex enhancement factor 

\be
\gamma_a(\infty) \simeq \frac{2h/R_a}{\alpha_1\ln(4h/R_a) - \alpha_2} \label{eq:FEF0}
\ee

\noi
where $\alpha_1,\alpha_2$ depend on the deviation from the linear line charge density. For the hemi-ellipsoid,
where the linear line charge density is applicable, $\alpha_1 = 1$ and $\alpha_2 = 2$.

The line charge model has also been successfully used to study shielding effects in a random array \cite{db_rudra}
of hemi-ellipsoidal emitters. It has been shown that when the mean separation of emitters is greater than $h/2$,
the apex field enhancement factor of the $i^{th}$  emitter is well approximated by 

\be
\gamma_a^{(i)}(\infty) \simeq \frac{2h/R_a}{\ln(4h/R_a) - 2 + \alpha_{S_i}}  \label{eq:FEF}
\ee

\noi
where $\alpha_{S_i}$ is a purely geometric  shielding term  due to all other emitters and depends on  quantities
such as the inter-emitter distances. Such a modular form is particularly useful since the shielding term $\alpha_S$ can
be calculated separately and plugged into the isolated emitter expression for AFEF.

Our aim here is to study the influence of anode proximity on the apex field enhancement factor
of an isolated curved emitter
using the line charge model. This has been investigated previously \cite{pogo2009} for a hemi-ellipsoid where the
line charge density $\Lambda(z)$ is linear ($\Lambda(z) = \lambda z$). We shall study the problem in a somewhat
different light such that it can be extended to other shapes where $\Lambda(z)$ is nonlinear. In the process, we shall
make certain approximations and arrive at a form 

\be
\gamma_a(D) \simeq \frac{2h/R_a}{\alpha_1 \ln(4h/R_a) - \alpha_2 - \alpha_A(D)}  \label{eq:FEF}
\ee

\noi
where $\alpha_A(D)$ is a geometric quantity that can be computed independently and plugged into the
expression for $\gamma_a(\infty)$ much in the same way as shielding expresses itself in the apex
field enhancement factor.

Apart from the local field at the apex, its variation close
to the apex is also crucial in determining the net field emission current. Recent
studies \cite{db_ultram,cosine} show that close to the emitter apex,

\be
\gamma(\infty,\ttl) = \gamma_a(\infty) \cos\ttl  \label{eq:coslaw}
\ee

\noi
where

\be
\cos\ttl = \frac{z/h}{\sqrt{(z/h)^2 + (\rho/R_a)^2}}  \label{eq:costtl}
\ee

\noi
for an axially symmetric emitter aligned along the asymptotic electric field $E_0 \hat{z}$,
when the anode is far away. The validity of the cosine law (Eq.~(\ref{eq:coslaw}))
when the anode is at a finite distane also needs to be examined.

In the following sections, we shall study the anode proximity effect
for a general line charge distribution $\Lambda(z)$
and in particular study  the effect of the anode on the field enhancement factor at the apex
and its neighbourhood. The results obtained, are verified numerically using
the finite element software COMSOL Multiphysics\textsuperscript{\textregistered} v5.4,
and subsequently used to understand scaling behaviour in the $I-V$ curves of an experimental situation.
In the concluding section, we shall also
discuss non-generic emitters where our results do not accurately predict the electrostatic field
behaviour.

\section{Linearly varying line charge distribution}

The linearly varying line charge distribution is a well-studied system. We shall revisit it here
and study the anode proximity effect at the emitter apex and its immediate neighbourhood since
field emission predominantly takes place from this region.

The potential at any point ($\rho,z$) due to a vertical line charge (along $z$-axis)  placed on a grounded conducting
plane in the presence of an electrostatic field $-E_0 \hat{z}$ can be expressed as

\be
\begin{split}
V(\rho,z) = & \frac{1}{4\pi\epsilon_0}\Bigg[ \int_0^L \frac{\Lambda(s)}{\big[\rho^2 + (z - s)^2\big]^{1/2}} ds ~
  - \\
  &  \int_0^L \frac{\Lambda(s)}{\big[\rho^2 + (z + s)^2\big]^{1/2}} ds \Bigg] + E_0 z \label{eq:pot}
\end{split}
\ee

\noi
where $L$ is the extent of the line charge distribution and $\Lambda(s) = \lambda s$ in the linear case.
The zero-potential contour
corresponds to the surface of the desired emitter shape so that
the parameters defining the line charge distribution
including its extent $L$, can, in principle be calculated by imposing the requirement that
the potential should vanish on the surface of the emitter.

Note that the potential at a height $D$ above the cathode, as given by Eq.~(\ref{eq:pot}),
is not $V_A = E_0 D$ as it should be if an anode is present at $z = D$ and held at a potential
$V_A$ while the cathode is grounded. The effect of the original line charge distribution
(and its image on the grounded cathode) on the anode can however be neutralized by placing a ``mirror''
line charge at a height $2D$ above the cathode. This however affects the potential on the cathode which
can again be neutralized by placing a ``mirror'' line charge (of the one at $z = 2D$) at $z=-2D$. The process
of correcting the potentials at the cathode and anode results in an infinite number of line charges whose
contribution can be summed to yield the corrected diode potential

\be
\begin{split}
  V_D&(\rho,z)  = E_0 z + \\
  & \frac{1}{4\pi\epsilon_0} \int_0^L ds \Bigg[ \frac{\Lambda(s)}{\sqrt{\rho^2 + (z - s)^2}} -
      \frac{\Lambda(s)}{\sqrt{\rho^2 + (z + s)^2}} +  \\
    &   \sum_{n=1}^\infty  \frac{\Lambda(s)}{\sqrt{\rho^2 + (2nD - z + s)^2}} -
         \frac{\Lambda(s)}{\sqrt{\rho^2 + (2nD - z - s)^2}} \\
  & +   \frac{\Lambda(s)}{\sqrt{\rho^2 + (2nD + z - s)^2}} - 
   \frac{\Lambda(s)}{\sqrt{\rho^2 + (2nD + z + s)^2}}  \Bigg]  \label{eq:potsum}
\end{split}
\ee

\begin{figure}[hbt]
  \begin{center}
    \vskip -4.75cm
\hspace*{-4.10cm}\includegraphics[scale=0.61,angle=90]{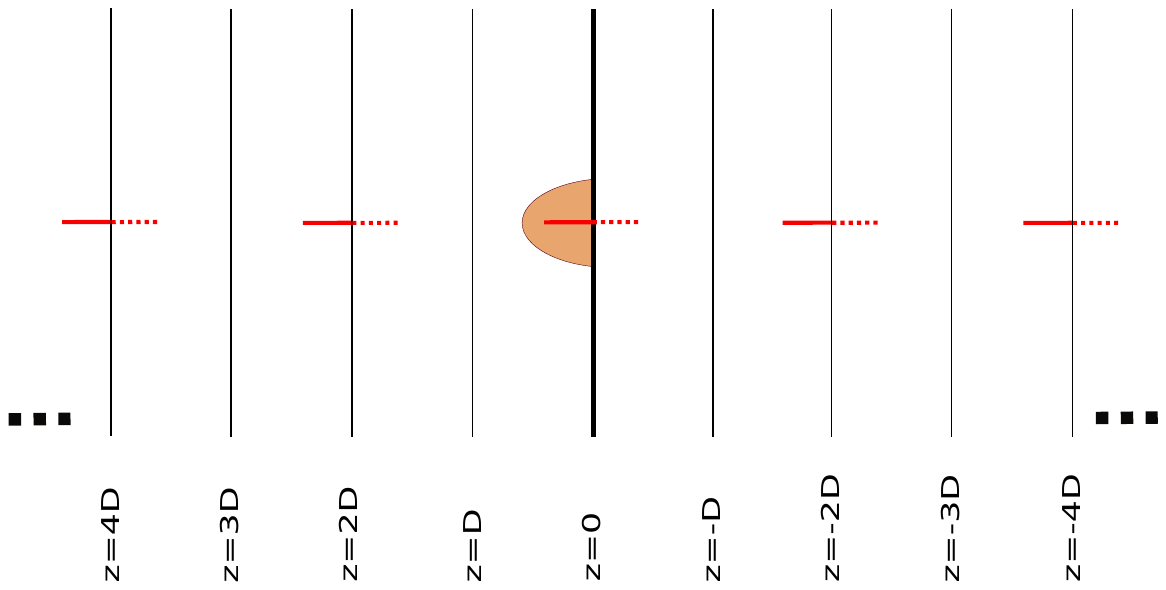}
\vskip -3.7 cm
\caption{The emitter on the cathode plane ($z = 0$) and the anode at a distance $D$ ($z = D$) can be modelled
using line charges and their successive images at the anode and cathode planes.}
\label{fig:error_gamma}
\end{center}
\end{figure}

For the linearly varying line charge density, $\Lambda(s) = \lambda s$, the value
of $\lambda$ can be determined by demanding that the potential vanish at the apex i.e. $V(\rho=0,z=h) = 0$.
Note that for the analytical results presented here, $ h \simeq L + R_a/2 > L$ as shown
elsewhere [\onlinecite{db_fef}]. Thus,

\be
\begin{split}
  -E_0 & h =  \frac{\lambda}{4\pi\epsilon_0}  \int_0^L ds \Bigg[ \frac{s}{h-s} - \frac{s}{h+s}
    + \sum_{n=1}^\infty \frac{s}{2nD + s - h} \\
    & - \frac{s}{2nD - h -s} + 
    \frac{s}{2nD - s + h} - \frac{s}{2nD + s + h} \Bigg] \label{eq:findlam}
\end{split}
\ee

\noi
so that

\be
\lambda = - \frac{4\pi\epsilon_0 E_0}{\ln[(h+L)/(h-L)] - 2L/h - \alpha_A}
\ee

\noi
where

\be
\begin{split}
\alpha_A = \frac{1}{h} \sum_{n=1}^\infty & 4nD \int_0^L  \Bigg[ \frac{s}{(2nD)^2 - (h+s)^2}  \\
  & -~\frac{s}{(2nD)^2 - (h-s)^2}\Bigg] ds.
\end{split}
\ee

\noi
Since $2nD > h+s$ for all $n$, 

\be
\begin{split}
  \alpha_A \simeq \frac{1}{h} \sum_{n=1}^\infty  \frac{4nD}{4n^2D^2} \int_0^L &  \Bigg[ s \Big\{ 1 + \big(\frac{h+s}{2nD}\big)^2\Big\} - \\
    & s \Big\{ 1 + \big(\frac{h-s}{2nD}\big)^2\Big\} \Bigg]  ds  \\
\end{split}
\ee

\noi
which simplifies as,  on using $L/D \simeq h/D$ simplifies as $h = L + R_a/2$ and $R_a << L$.

\be
\begin{split}
  \alpha_A^{(1)} & \simeq \frac{1}{4D^3} \sum_{n=1}^{\infty} \frac{1}{n^3} \int_0^L 4s^2 ds \\
  & = \frac{1}{3} \Big(\frac{L}{D}\Big)^3 \sum_{n=1}^{\infty} \frac{1}{n^3} \\
  & \simeq \frac{1}{3} \Big(\frac{h}{D}\Big)^3 \zeta(3)  \label{eq:alpha_s1}
\end{split}
\ee

\noi
where $\zeta(3) \simeq 1.2020569$. In the above, we have used $L/D \simeq h/D$ since $h = L + R_a/2$ and $R_a << L$.
The quantity $\alpha_A^{(1)}$ is a first order approximation of the exact result
which can  be arrived by writing the
anode shielding factor as

\be
\begin{split}
  \alpha_A = & \frac{1}{h} \sum_{n=1}^\infty \int_0^L  \Bigg[ \Big\{\frac{s}{2nD-h+s} - \frac{s}{2nD-h-s} \Big\} \\
    & ~~~~- \Big\{\frac{s}{2nD+h+s} - \frac{s}{2nD+h-s} \Big\} \Bigg]~ds \\
\end{split}
\ee

\noi
which on integration yields

\be
\begin{split}
\alpha_A  = & \frac{2}{h} \sum_{n=1}^\infty \Bigg[ (2nD-h)  \tanh^{-1}\frac{L}{2nD-h} \\
  & ~~~~~~~~~~ - (2nD+h) \tanh^{-1}\frac{L}{2nD+h} \Bigg]. \label{eq:tanh}
\end{split}
\ee

\noi
Eq.~\ref{eq:tanh} is the exact result for $\alpha_A$ which in turn can be used to obtain
the first and second approximations as follows.
On expanding $\tanh^{-1}(z)$ in its Maclaurin series

\be
\tanh^{-1}(z) = z + \frac{z^3}{3} + \frac{z^5}{5} + \ldots
\ee

\noi
and keeping the first 3 terms, the leading terms in powers of  $(h/D)$ are

\be
\begin{split}
\alpha_A^{(2)} & \simeq  \frac{1}{3} \Big(\frac{L}{D}\Big)^3  \sum_{n=1}^\infty \frac{1}{n^3} \frac{1}{\big[1 - h^2/(2nD)^2\big]^2} \\
  & ~~~~~~~~~ + \frac{1}{10} \Big(\frac{L}{D}\Big)^5 \sum_{n=1}^\infty \frac{1}{n^5} \frac{1}{\big[1 - h^2/(2nD)^2\big]^4} \\
& \simeq  \Big(\frac{h}{D}\Big)^3 \Bigg[ \frac{\zeta(3)}{3} + \frac{4}{15}\Big(\frac{h}{D}\Big)^2 \zeta(5) + {\cal{O}}\Big(h^4/D^4\Big) \Bigg] \label{eq:alphaA}
\end{split}
\ee

\noi
where, in the last step, we have used $h/2D < 1$ to expand the expressions in the denominator and substituted $L/D$ with $h/D$.
Note that $\zeta(5) = \sum_{n=1}^\infty 1/n^5 \simeq 1.03692$. It is adequate to retain the first two terms
in a calculation of the enhancement factor for most plate separations $D$. However, if $D$ is only slightly larger
than $h$ as in case of certain applications, the summation in Eq.~(\ref{eq:tanh}) should be evaluated numerically for
accurate results. 

\begin{figure}[hbt]
  \begin{center}
    \vskip -0.5cm
\hspace*{-1.0cm}\includegraphics[scale=0.35,angle=0]{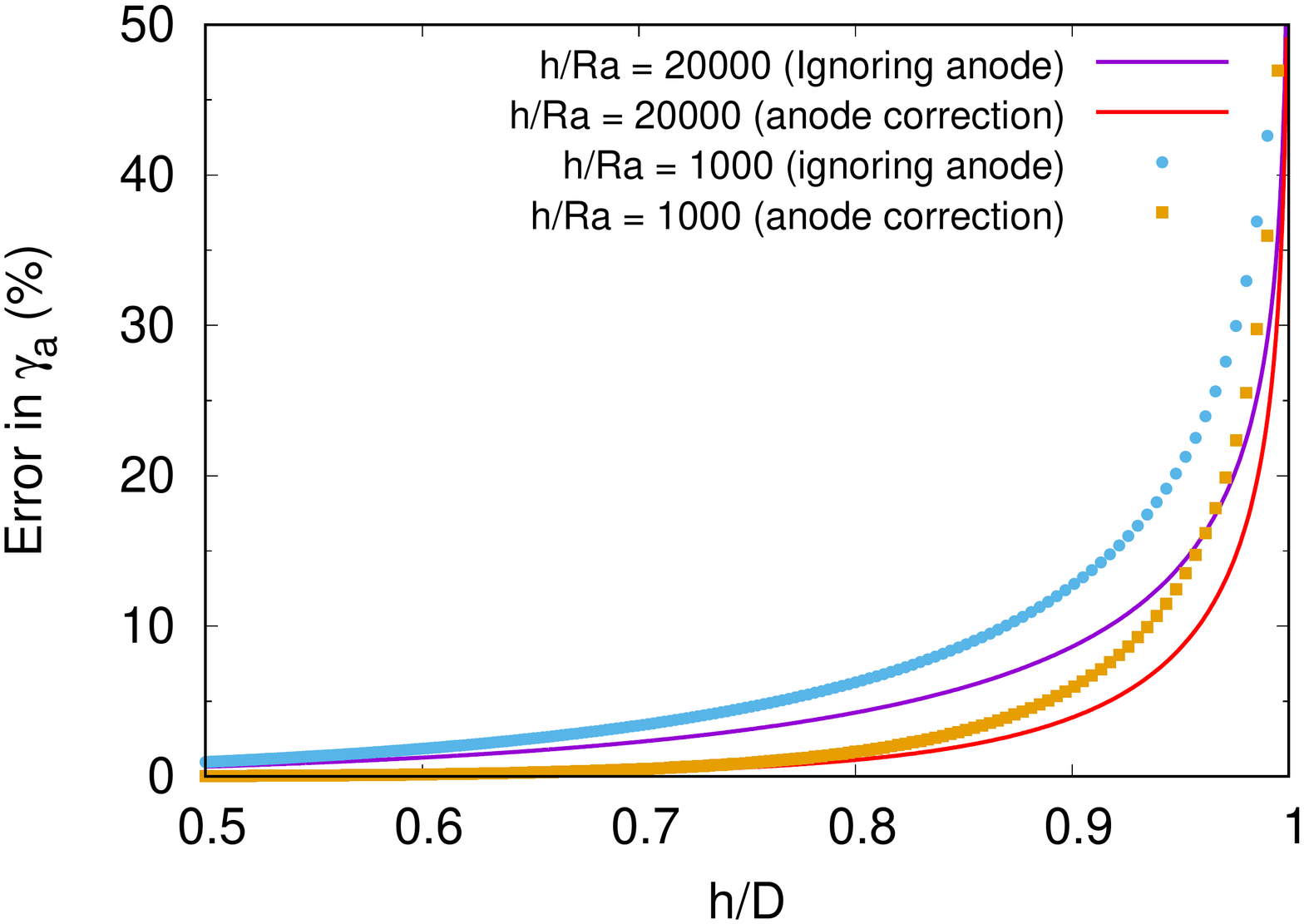}
\caption{The error in $\gamma_a$ without and with anode-correction using the second-order expression for $\alpha_A$ (Eq.~\ref{eq:alphaA}) as compared to the result for $\gamma_a(D)$ using the exact expression for
  $\alpha_A$ (Eq.~\ref{eq:tanh}). The cases
  studied are (i) $h/R_a = 20000$ (solid lines) with the bottom curve incorporating
  anode correction using Eqns.~(\ref{eq:withalpha}) and Eq.~(\ref{eq:alphaA}) and  (ii) $h/R_a = 1000$ with (solid squares) and without (solid circles)
  anode correction.}
\label{fig:error_gamma}
\end{center}
\end{figure}

It now remains to determine $E_z$ at the apex ($\rho = 0, z = h$). On differentiating Eq.~(\ref{eq:potsum})
with respect to $z$ and substituting for the value of $\rho$ and $z$, we have

\be
\begin{split}
  \frac{\partial V}{\partial z} & \Big|_{\rho=0,z=h}  = \frac{\lambda}{4\pi\epsilon_0}  \int_0^L ds \Bigg[ \frac{s}{(h+s)^2} -
    \frac{s}{(h-s)^2} + \\
    & \sum_{n=1}^\infty \Big\{ \frac{s}{(2nD - h + s)^2} - \frac{s}{(2nD - h -s)^2} \\
    & + \frac{s}{(2nD + h + s)^2} - \frac{s}{(2nD + h -s)^2} \Big\} \Bigg] + E_0
\end{split}
\ee

\noi
which on integration yields

\be
\begin{split}
  \frac{\partial V}{\partial z}& \Big|_{\rho=0,z=h} = E_0 + \frac{\lambda}{4\pi\epsilon_0} \Bigg[ \Big\{ -\frac{2hL}{h^2 - L^2} +
  \ln\frac{h+L}{h-L} \Big\} \\
  & + \sum_{n=1}^\infty \Big\{-\frac{2(2nD - h)L}{(2nD - h)^2 - L^2} + \ln\frac{2nD - h + L}{2nD - h - L} \Big\} \\
  & +  \sum_{n=1}^\infty \Big\{-\frac{2(2nD + h)L}{(2nD + h)^2 - L^2} + \ln\frac{2nD + h + L}{2nD + h - L} \Big\} \Bigg]
\end{split}
\ee

\noi
Since $h - L \simeq R_a/2$, the term in the first curly bracket dominates for a sharp emitter and all the other terms can be
neglected. Thus at the apex,

\bea
E_a & = & -\frac{\partial V}{\partial z} \Big|_{\rho=0,z=h}  \nonumber \\
& \simeq &  \frac{\lambda}{4\pi\epsilon_0} \frac{2hL}{h^2 - L^2} \nonumber \\
& = & -E_0~ \frac{2hL/(h^2 - L^2)}{\ln[(h+L)/(h-L)] - 2L/h - \alpha_A} \nonumber  \\
& \simeq & -E_0~ \frac{2h/R_a}{\ln(4h/R_a) - 2 - \alpha_A} = -\gamma_a(D) E_0  \label{eq:withalpha}
\eea

\noi
where $hL \simeq h^2$, $h + L \simeq 2h$ and $L/h \simeq 1$.
Alternately,

\bea
\gamma_{a}(D) &  = & \frac{\gamma_{a}(\infty)}{\Big[1 - \alpha_A/(\ln(4h/R_a) - 2)  \Big]} \label{eq:withalpha1} \\
& \simeq & \gamma_{a}(\infty) \Big[ 1 + \alpha_A/\{\ln(4h/R_a) - 2\} \Big]
\eea

\noi
where the last line holds for a sharp emitter and  

\be
\gamma_a(\infty) = \frac{2h/R_a}{\ln(4h/R_a) - 2}. \label{eq:gaminf0}
\ee

\noi
Thus, the correction term depends on the apex radius of curvature.

Fig.~\ref{fig:error_gamma} shows a plot of error in apex field enhancement factor
when the anode is placed at a distance $D$ and the ratio $h/D$ is varied.
The apex enhancement factor is computed using Eq.~(\ref{eq:gaminf0}) when the anode is considered to be at infinity
and Eq.~(\ref{eq:withalpha1}) along with Eq.~(\ref{eq:alphaA}) when the second order anode correction is used.
These are compared with $\gamma_a(D)$ computed using
Eq.~(\ref{eq:withalpha1}) and Eq.~(\ref{eq:tanh}) (considered exact here) in order to find the errors.
Two cases are considered: $h/R_a = 20000$ and $h/R_a = 1000$. In the first case (solid lines), the
error is below $5\%$ for $h/D < 0.9$ while in the second case, this happens for a somewhat larger $D$.

\begin{figure}[hbt]
  \begin{center}
    \vskip -0.5cm
\hspace*{-1.0cm}\includegraphics[scale=0.35,angle=0]{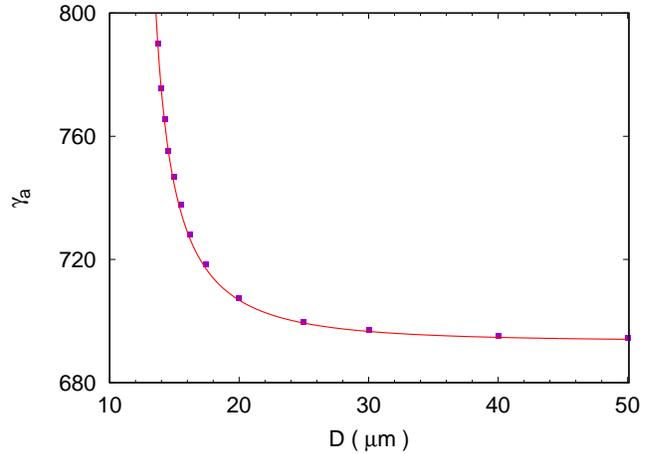}
\vskip -0.70 cm
\caption{The apex field enhancement factor $\gamma_a$ is plotted against the anode-cathode distance $D$.
  The solid squares are computed using COMSOL while the continuous curve is calculated
  using Eq.~(\ref{eq:withalpha}) with $\alpha_A$ given by Eq.~(\ref{eq:tanh}). The height
  of the hemiellipsoid is $h = 12.5\mu$m.}
\label{fig:ellip}
\end{center}
\end{figure}

We next test the accuracy of Eq.~(\ref{eq:withalpha1}) with $\alpha_A$ given by Eq.~(\ref{eq:tanh}).
For this purpose the finite element software COMSOL v5.4 is used to model a grounded  hemiellipsoid on a planar
cathode with the (planar) anode at a distance $D$ and maintained at a potential $V_A$.
The height of the ellipsoid is $h = 12.5\mu$m while the apex radius of curvature is $R_a = 5$nm. The
base radius is thus $b = 250$nm. The distance $D$ is varied and the apex field enhancement factor
is recorded. The results are shown in fig.~\ref{fig:ellip}. Clearly, Eqns.~(\ref{eq:withalpha}) and (\ref{eq:tanh})
reproduce the results quite accurately.

\section{Nonlinear line charge distribution}

\subsection{Analytical derivation}

For a  nonlinear line charge distribution $\Lambda(s) = sf(s)$, the expression for
$-E_0 h$ in Eq.~(\ref{eq:findlam}) modifies as

\be
\begin{split}
   & \frac{1}{4\pi\epsilon_0}  \int_0^L ds f(s) \Bigg[ \frac{s}{h-s} - \frac{s}{h+s}
    + \sum_{n=1}^\infty \frac{s}{2nD + s - h} \\
    & - \frac{s}{2nD - h -s} + 
    \frac{s}{2nD - s + h} - \frac{s}{2nD + s + h} \Bigg] \label{eq:findlam1}
\end{split}
\ee

\noi
which can be integrated by parts to yield

\be
\begin{split}
  -E_0 h & =  \frac{f(L)}{4\pi\epsilon_0} \Bigg[ h \ln\Big(\frac{h+L}{h-L}\Big)  (1 - \cc_1) - 2L (1 - \cc_2)  \\
    & ~~~~~~~~~~~~~ - h ~\alpha_A (1 - \cc)  \Bigg] 
\end{split}
\ee

\noi
where

\be
\cc  =  \int_0^L \frac{f'(s)}{f(L)} \frac{ \cd_-\tanh^{-1}\Big(\frac{s}{\cd_-}\Big) - \cd_+\tanh^{-1}\Big(\frac{s}{\cd_+}\Big)}{\cd_-\tanh^{-1}\Big(\frac{L}{\cd_-}\Big) - \cd_+\tanh^{-1}\Big(\frac{L}{\cd_+}\Big) }, \\
\ee

\noi
$\cd_+ = 2nD + h$, $\cd_- = 2nD - h$ and

\bea
\cc_1 & = & \int_0^L ~ds~ \frac{f'(s)}{f(L)}\frac{\ln\Big(\frac{h+s}{h-s}\Big)}{\ln\Big(\frac{h+L}{h-L}\Big)} ds \\
\cc_2 & = &  \int_0^L~ds~ \frac{f'(s)}{f(L)} \frac{s}{L}  ds
\eea

\noi
Thus,

\be
f(L) = \frac{4\pi\epsilon_0 E_0}{\ln\Big(\frac{h+L}{h-L}\Big)  (1 - \cc_1) - 2\frac{L}{h}(1 - \cc_2) - \alpha_A ( 1 - \cc )} \label{eq:FL}
\ee

\noi
where $\alpha_A$ is given by Eq.~\ref{eq:tanh}. While the quantities $\cc_1$ and $\cc_2$ must be considered for
any departure from the ellipsoidal shape corresponding to a linear line charge density, 
$\cc$ in general is a small quantity that we must eventually
neglect in order to get a useful form for the apex field enhancement factor $\gamma_a(D)$.
Note that as in the linear case, a Maclaurin series expansion of $\tanh^{-1}$ can be used to approximate
$\alpha_A$ when the anode is not too close to the apex.

Using methods similar to that for linear systems, the local field at the apex is expressed as

\bea
E_z & \simeq & \frac{f(L)}{4\pi\epsilon_0} \frac{2hL}{h^2-L^2}( 1 - \cc_0)
\eea

\noi
where

\be
\cc_0 =  \int_0^L \frac{f'(s)}{f(L)} \frac{s/(h^2 - s^2)}{L/(h^2 - L^2)}  ds
\ee

\noi
is small for a sharp emitter \cite{db_fef} but nevertheless can be retained for completeness.
Finally, as shown in [\onlinecite{db_fef}], $R_a = (h^2 - L^2)/h$ so that
with the approximations already introduced ($h = L + R_a/2$ with $R_a << L$)
\be
  \gamma_a(D)   =  \frac{(1 - \cc_0) 2h/R_a}{(1 - \cc_1)\ln(\frac{4h}{R_a}) - 2(1 - \cc_2) - \alpha_A (1 - \cc)}  \\
  \label{eq:gamma_anode}
\ee

\noi
while 

\be
\gamma_a(\infty) = \frac{(1 - \cc_0) 2h/R_a}{(1 - \cc_1)\ln(4h/R_a) - 2(1 - \cc_2)}. \label{eq:gaminf}
\ee

\noi
Eq.~(\ref{eq:gamma_anode}) is the central result connecting the local field enhancement and
anode proximity. Note that $\cc,\cc_0,\cc_1,\cc_2$ are zero for a hemi-ellipsoid since $f'(s) = 0$.

The presence of the anode thus leads to an increase in the local field at the emitter apex for all
smooth  emitters irrespective of its shape since $\alpha_A > 0$. The quantities $\cc_1$ and $\cc_2$ 
are in general nonzero even for sharp nonlinear emitters though the upper bound on $\cc_1$ can be shown to
vanish weakly as $h/R_a$ tends to infinity. The quantities $\cc_0$ and $\cc$ are in general small
as our numerical investigations reveal.

Two approximations lead to a particularly useful formula for $\gamma_a(D)$. Since the nonlinear correction terms
depend on the line charge density, the quantities $\cc, \cc_0, \cc_1$ and $\cc_2$ must in principle
depend on $D$. We shall however assume that these are weakly dependent on $D$ and can be assumed to
be constants as a first approximation. Further, we shall neglect $\cc$ altogether. Thus, $\gamma_a(D)$
can be determined if $\gamma_a(\infty)$ is known since $\alpha_A$ is a purely geometric quantity.

\subsection{Numerical Verification}
The results presented in the preceding sub-section hold for the anode at a finite distance ($D$) from the cathode plane.
These have been derived using the nonlinear line charge model by ensuring that the presence of the line charge does
not alter the potential at the anode and cathode planes as well as the emitter apex. As in the linear case,
the derivation necessitated certain
approximations such as $h >> R_a$ and the assumption that shape distortion of the zero potential (emitter) surface does not occur
due to the presence of the anode. We shall now verify the results numerically.

\begin{figure}[h]
  \begin{center}
    \vskip -0.9cm
\hspace*{-1.0cm}\includegraphics[scale=0.32,angle=0]{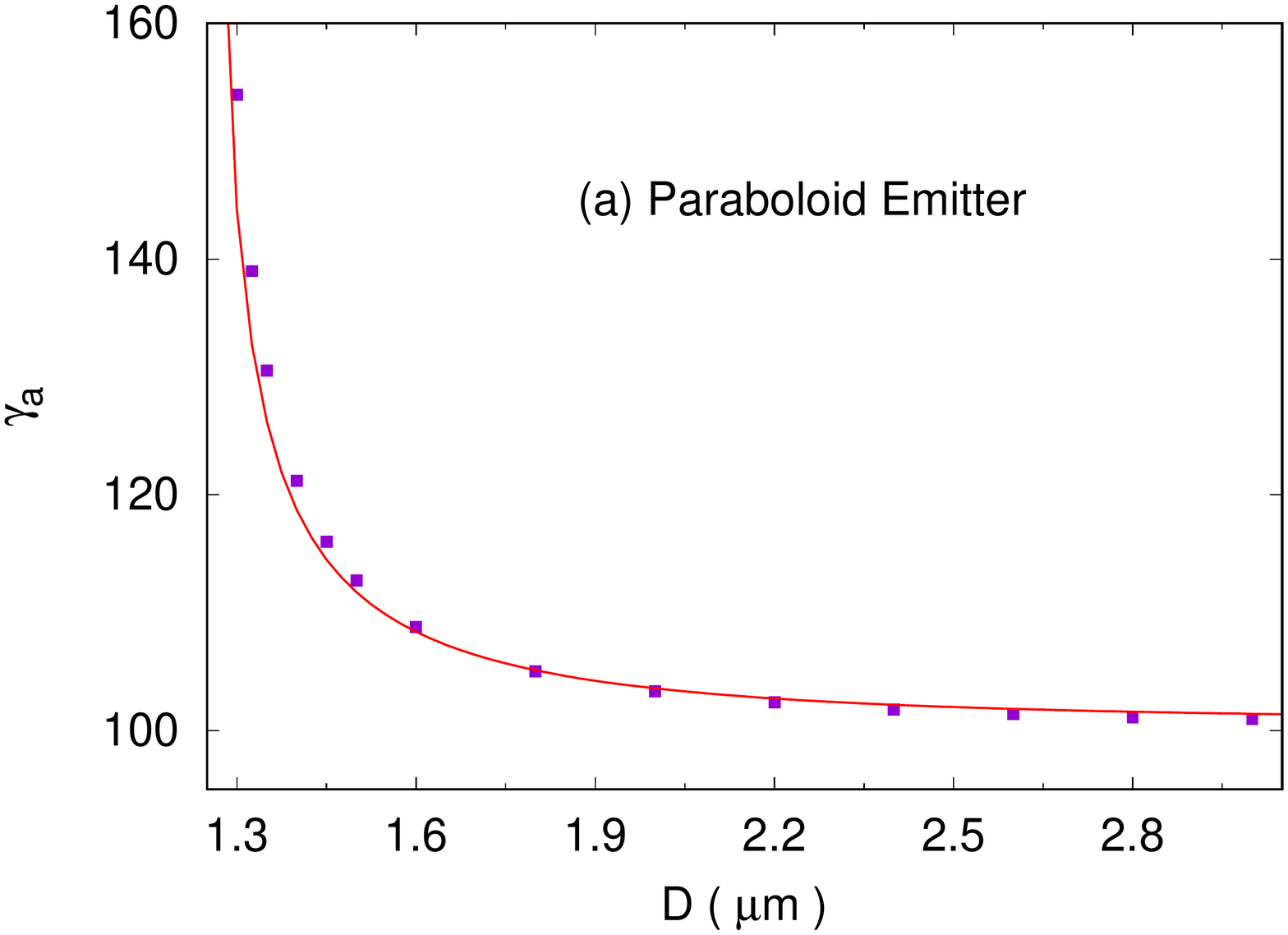}
\vskip -0.90 cm
\hspace*{-1.0cm}\includegraphics[scale=0.32,angle=0]{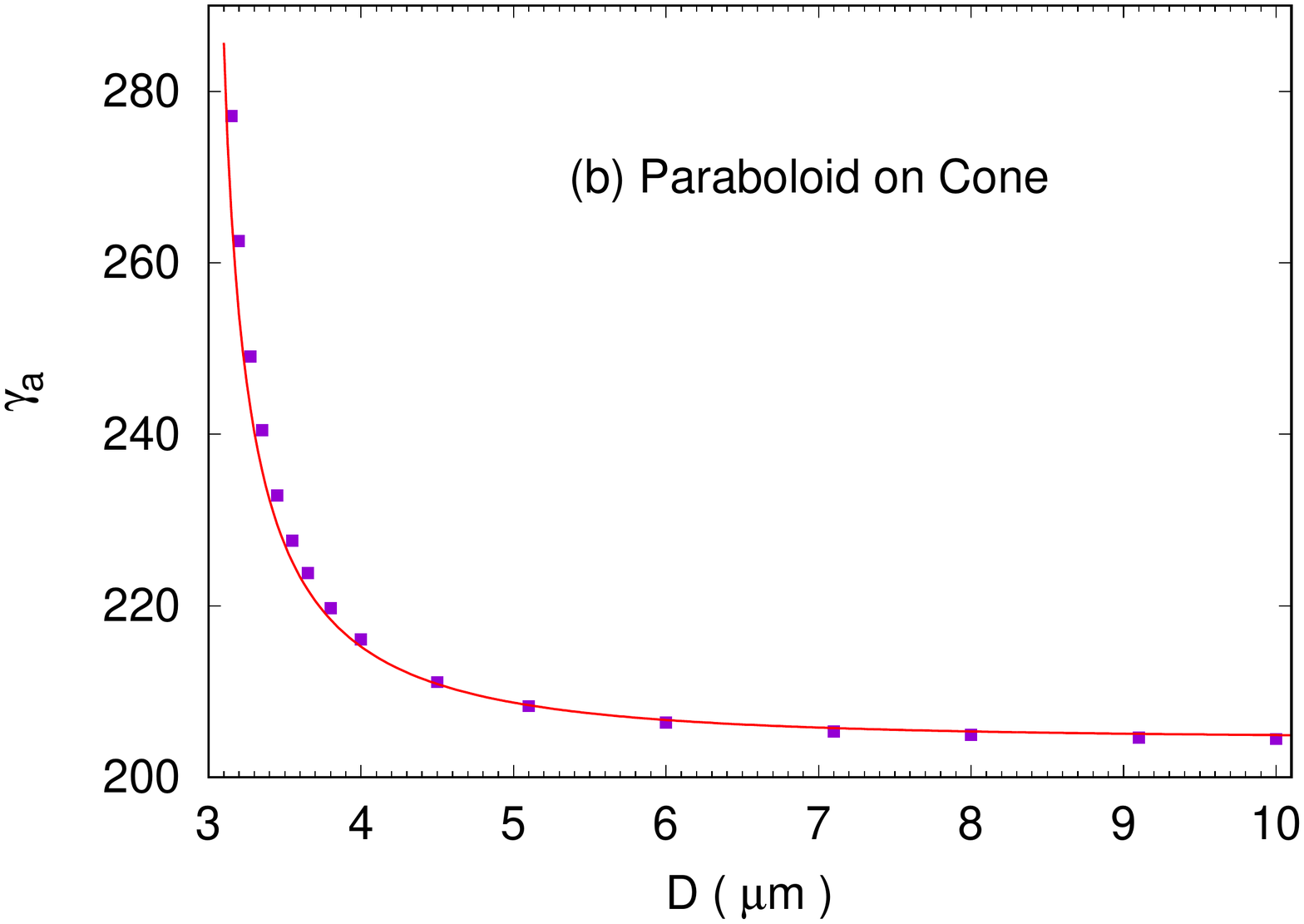}
\vskip -0.90 cm
\hspace*{-1.0cm}\includegraphics[scale=0.32,angle=0]{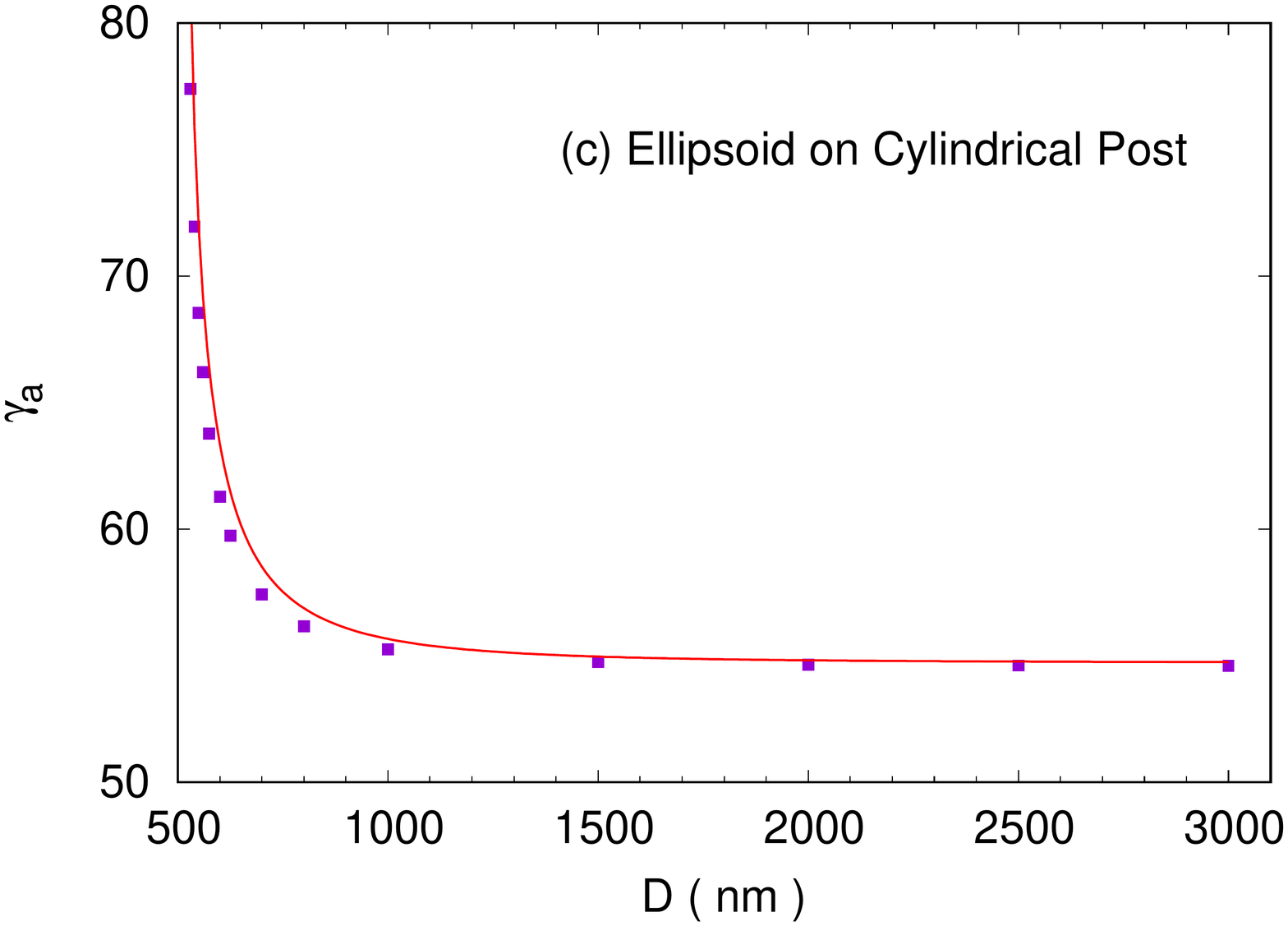}
\vskip -0.90 cm
\caption{The apex field enhancement factor plotted against the anode-cathode distance $D$ for
  (a) a paraboloid (b) a rounded cone and (c) ellipsoid on a cylindrical post. The solid squares are
obtained directly from COMSOL.}
\label{fig:nonlin}
\end{center}
\end{figure}

The curved emitter in a planar diode configuration has been modeled using the finite element software COMSOL v5.4. The shapes
chosen are  (a) a paraboloid  (b) a paraboloidal cone and (c) an ellipsoid on a cylindrical post (ECP).
The process of verification involves computation of the field at the apex and along the surface of the emitter using COMSOL.
The field at the apex immediately yields the apex field enhancement factor $\gamma_a(D)$ while the (normal) field along the surface
is used to first compute the surface charge density. For the axially ($z$) symmetric emitters, the surface charge density
$\sigma(z) = \epsilon_0 E(z)$ where $E(z)$ is the normal field at the point $(\rho,z)$ on the surface $\rho = \rho(z)$.
The surface charge density can then be projected along the axis ($z$) to determine the line charge density \cite{jap16}

\be
\Lambda(z) = 2\pi \rho(z) \sqrt{1 + \big(\frac{d\rho}{dz}\big)^2}~ \sigma(z)
\ee

\noi
at each point $z < h$. The extent of the line charge $L$ is then fixed by demanding that the potential be zero
at the apex. The (nonlinear) line charge density so obtained can then be used to compute the quantities
$(1~-~\cc_k)$, $k = 0,1,2$ using the surface electric field for large $D$. Thus, $\gamma_a(\infty)$ can be
calculated using Eq.~\ref{eq:gaminf}. The expression for $\alpha_A$ can be computed using Eq.~\ref{eq:tanh}
and plugged into the expression for $\gamma_a(D)$ in Eq.~\ref{eq:gamma_anode}. The quantity
$\cc$ is found to be small in all cases and is  neglected in the numerical results presented in Fig.~\ref{fig:nonlin}.

The shapes considered are (a) a paraboloid of height $1.25\mu$m and apex radius $5$nm (b) a conical base of half angle
$5^\circ$ topped by a paraboid. The  total emitter height is $3\mu$m and the apex radius of curvature is $5$nm (c) a cylinder
of height 450nm with an ellipsoid cap of height 50nm and apex radius 5nm. In each of the three cases, the
quantities $1-\cc_0$, $1-\cc_1$ and $1-\cc_2$ have been evaluated at $D > 4h$ so as to mimic the anode at infinity.
These have been used in Eq.~\ref{eq:gamma_anode} together with $\alpha_A$ evaluated using Eq.~\ref{eq:tanh}. The
results thus obtained are represented by continuous curve in Fig.~\ref{fig:nonlin} while the solid squares have been evaluated
using the COMSOL data at the emitter apex. The close agreement suggests that the approximations used are
justified and Eq.~\ref{eq:gamma_anode} with $1-\cc \simeq 1$ can be used to get the apex field enhancement when the
anode is in close proximity.

\section{Anode proximity and the enhancement factor variation near the apex}

We have so far dealt with the effect of anode on the apex field enhancement factor. In order to
evaluate the net field emission current, it is important to know about the variation of
local field in the neighbourhood of the apex. When the anode is sufficiently far away, it has
been established recently that $\gamma(\infty,\ttl) = \gamma_a(\infty)\cos\ttl$. Our interest
here is to determine the enhancement factor variation $\gamma(D,\ttl)$ when the anode is at a
finite distance from the cathode.

For a general nonlinear line charge distribution for a sharp emitter ($h/R_a$ large), the electric field
components for small $\rho$ can be expressed as

\be
\begin{split}
E_\rho = & -\frac{\partial V}{\partial \rho}  = -\frac{\rho}{4\pi\epsilon_0} \Bigg[ \int_0^L \frac{s f(s)}{[\rho^2 + (z+s)^2]^{3/2}}ds - \\
      & \int_0^L \frac{s f(s)}{[\rho^2 + (z-s)^2]^{3/2}}ds  + \sum_{n=1}^\infty T_1 + T_2 + T_3 + T_4 \Bigg] 
\end{split}
\ee

\noi
where

\bea
T_1 & = & -\int_0^L \frac{s f(s)}{[\rho^2 + (2nD - z + s)^2]^{3/2}}ds \\
T_2 & = & ~\int_0^L \frac{s f(s)}{[\rho^2 + (2nD - z - s)^2]^{3/2}}ds \\
T_3 & = & ~\int_0^L \frac{s f(s)}{[\rho^2 + (2nD + z + s)^2]^{3/2}}ds \\
T_4 & = & -\int_0^L \frac{s f(s)}{[\rho^2 + (2nD + z - s)^2]^{3/2}}ds
\eea

\noi
For $\rho$ small ($ < R_a/2$) and $D > 2h$, each integrand can be expanded to extract the leading term in $\rho$. For example

\be
\int_0^L \frac{s f(s)}{[\rho^2 + (z \pm s)^2]^{3/2}}ds = \int_0^L \frac{s f(s)}{(z \pm s)^3} \Big[1 - \frac{3}{2} \frac{\rho^2}{(z \pm s)^2} \Big]~ds
\ee

\noi
which on combining and keeping the leading term, yields  

\be
\begin{split}
 \int_0^L   \Bigg[ & \frac{s f(s)}{(z+s)^3} -  \frac{s f(s)}{(z-s)^3} \Bigg] ds \\
 = & - f(L) \frac{2L^3}{(z^2 - L^2)^2} \Bigg[ 1 - \int_0^L ds~ \frac{f'(s)}{f(L)} \frac{s^3/(z^2 - s^2)^2}{L^3/(z^2 - L^2)^2} \Bigg]  \\
\simeq  &  -f(L) \frac{2L^3}{(z^2 - L^2)^2}  \Big[1 - \cc_0 \Big]
\end{split}
\ee

\noi
Note that $\cc_0$ is vanishing small for sharp emitters and can be neglected.
The other two pairs of integrands can be similarly combined to yield

\bea
T_1 + T_2 & \sim & \frac{1}{(2nD - z)^2 - L^2} \\
T_3 + T_4 & \sim & \frac{1}{(2nD + z)^2 - L^2},
\eea

\noi
none of which contribute as significantly for a sharp emitter. Thus,

\be
E_\rho = \frac{f(L)}{4\pi\epsilon_0} \frac{2L^2}{(z^2 - L^2)} \frac{L \rho}{z^2 - L^2} + {\cal{O}}(\rho^3).
\ee

\noi
This is identical to the result when the anode is at infinity except that $f(L)$ is now
as given in Eq.~(\ref{eq:FL}).

The calculation of $E_z$ proceeds along lines similar to the anode-at-infinity case \cite{cosine} so
that

\be
E_z \simeq \frac{f(L)}{4\pi\epsilon_0} \frac{2zL}{z^2 - L^2} \Bigg[ 1 -
  \frac{\rho^2}{2} \frac{4L^2}{(z^2 - L^2)^2} \Bigg]. \label{eq:Ezfin}
\ee

\noi
with $f(L)$ given by Eq.~(\ref{eq:FL}).

Now, consider the point $(\rho,z)$ to be located on the surface of a axially symmetric emitter aligned along the
$z$-axis, near the apex
with $\rho$ and $z$ related by $z \simeq h - \frac{\rho^2}{2R_a}$. The electric field lines
are normal to this parabolic surface and thus in the direction

\be
\hat{n} = \frac{1}{\sqrt{1 + (\rho/R_a)^2}} (\frac{\rho}{R_a} \hat{\rho},\hat{z}).
\ee

It thus follows using $\sqrt{E_\rho^2 + E_z^2}$ or $\vec{E}.\hat{n}$ and Eq.~(\ref{eq:FL}) for $f(L)$ that

\be
|E(\rho,z)| = E_0 \gamma(D,\ttl) = E_0 \gamma_a(D) \cos\ttl.  \label{eq:gencos}
\ee

\noi
Thus, the generalized cosine law of enhancement factor variation near the apex is unaffected by the presence of the
anode even though the local field itself increases in magnitude.

\begin{figure}[hbt]
  \begin{center}
    \vskip -0.5cm
\hspace*{-1.0cm}\includegraphics[scale=0.35,angle=0]{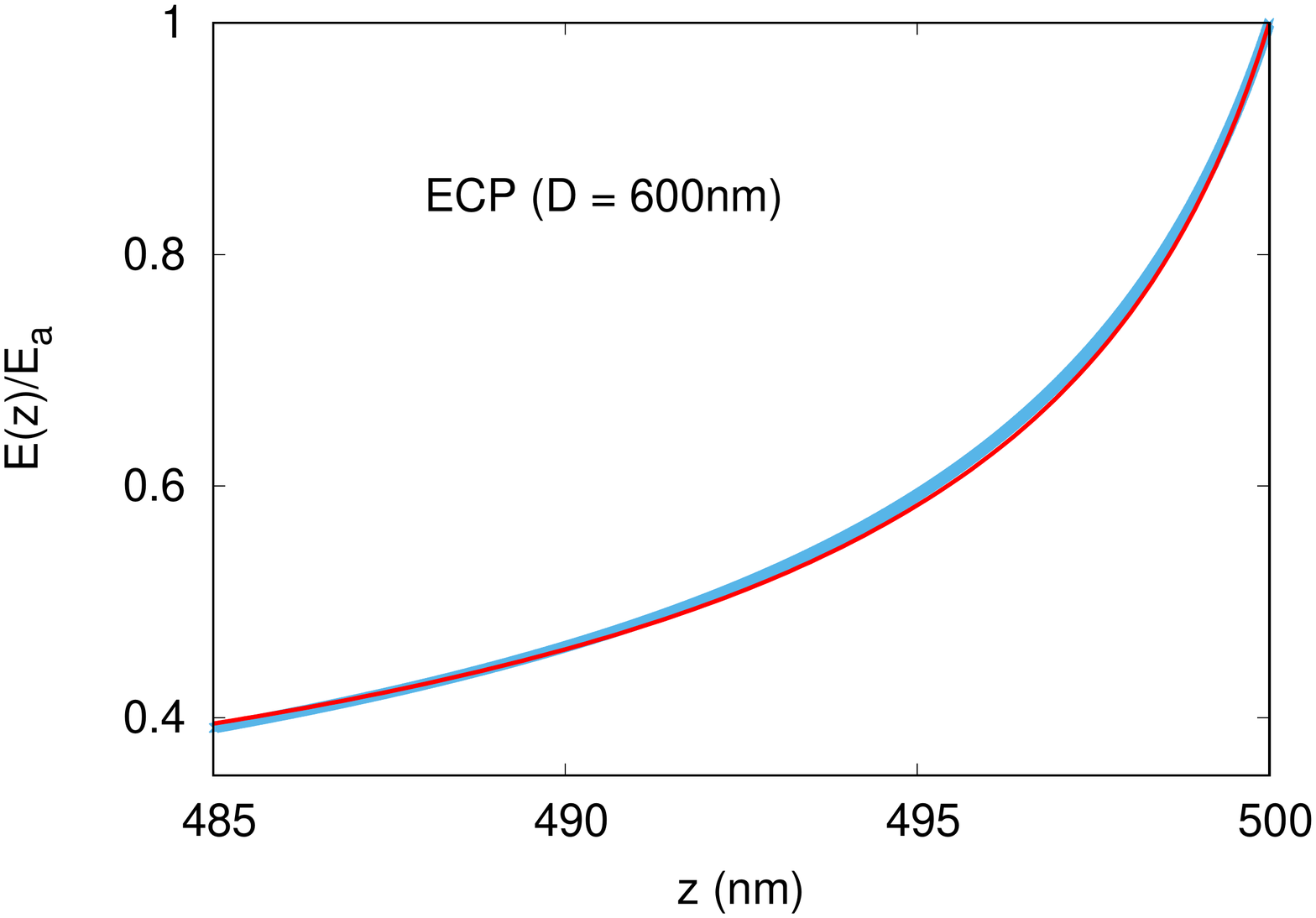}
\vskip -0.70 cm
\caption{Variation of the apex field enhancement $\gamma_a$ on the surface of an ellipsoid-on-cylindrical-post (ECP) emitter close to the
  apex. The apex radius is 5nm and its height is 500nm. The quantity plotted is $E(z)/E_a$ where $E_a = E(h)$ is the field at the apex while $E(z)$ is the
  field at a height $z$ from the cathode plane. The thin continuous curve is $\cos\ttl$ while
  the points (broad curve) is computed using COMSOL data. }
\label{fig:costtl}
\end{center}
\end{figure}

In order to verify this numerically, we shall consider an ellipsoid on a cylindrical post (ECP) with $h = 500$nm
and $R_a = 5$nm as discussed before. The anode is placed close  to the emitter apex tip at $D = 600$nm.
The comparison between Eq.~\ref{eq:costtl} and COMSOL data is shown in Fig.~\ref{fig:costtl}. The close
agreement shows that the variation of the surface electric field close to the apex is governed by the
generalized cosine law even when the anode is in close proximity.

\section{Validation using an experimental result}

An experiment carried out by Cabrera et al \cite{cabrera} reported scaling behaviour
in the $I-V$ graph of a nano-diode. The emitter is a
rounded conical structure of length $250\mu$m, having apex radius $R_a$ in the range 5-30nm, mounted on a cylindrical
base of length $1750\mu$m. The total emitter height $h = 2000\mu$m.
The movable anode is a planar electrode at a distance $d$ from
the apex. The distance $d$ is varied from a few nanometer to a few millimeter.
In particular ranges of $d$, the current was found to scale as $I \sim V d^{-\lambda}$
so that all the I-V curves in a given range collapse onto a single curve on scaling the applied voltage by
$R(d) \sim d^{-\lambda}$.

The argument put forward to support the scaling behaviour, rests on the observation that, a constant current
on scaling implies an identical tunneling potential at the apex and in its neighbourhood. Thus the key to the explanation
must lie in the behaviour of the local field in the apex-neighbourhood as $d$ is varied.

We have shown in the previous section that the local field variation in the apex neighbourhood obeys the generalized cosine law
$E(\rho,z) = E_0 \gamma_a(D) \cos\ttl$ where $\cos\ttl = (z/h)/\sqrt{(z/h)^2 + (\rho/R_a)^2}$. Also, since
$D = h + d$, the only term dependent on $d$ is the local field at the apex, $E_0 \gamma_a(D)$. We thus need
to study the scaling behaviour of the enhancement factor

\be
\gamma_a(D)  \simeq \frac{(1 - \cc_0) 2h/R_a}{(1 - \cc_1)\ln(4h/R_a) - (1 - \cc_2)2 - (1 - \cc)\alpha_A}.
\ee

\noi
The above equation can be expressed as

\be
\gamma_a(D) \simeq \frac{2h/R_a}{\alpha_1 \ln(4h/R_a) - \alpha_2 - \alpha_A (1 - \cc)/(1 - \cc_0)}
\ee

\noi
where $\alpha_1 = (1 - \cc_1)/(1 - \cc_0)$ and $\alpha_2 = 2(1 - \cc_2)/(1 - \cc_0)$.
As mentioned earlier, both $\cc_0$ and $\cc$ are small for generic emitters with $\cc$ slightly smaller than $\cc_0$. 
We shall therefore use the approximation $(1 - \cc)/(1 - \cc_0) \simeq 1$.
Thus,

\bea
\gamma_a(D) & = & \frac{\gamma_a(\infty)}{1 - \alpha_A/(\alpha_1 \ln(4h/R_a) - \alpha_2)} \\
& = & \frac{\gamma_a(\infty)}{1 - (\alpha_A \gamma_a(\infty))/(2h/R_a)} \label{eq:approxgam}
\eea

\noi
where $\gamma_a(\infty) = (2h/R_a)(\alpha_1 \ln(4h/R_a) - \alpha_2)$.
Thus, if the enhancement factor for the anode at a large distance is known, $\gamma_a(D)$ can
be determined using the height $h$, the apex radius of curvature $R_a$ and $\alpha_A$ with

\be
\begin{split}
  \alpha_A = & \frac{2}{h} \sum_{n=1}^\infty \Bigg[ (2nD-h)  \tanh^{-1}\frac{L}{2nD-h} \\
    & ~~~~~~~~~~ - (2nD+h) \tanh^{-1}\frac{L}{2nD+h} \Bigg] \label{eq:tanh1}.
\end{split}
\ee

The apex field enhancement factor for the anode at a large distance can be determined using
COMSOL. The emitter is modelled as a parabolic cone of net height $250\mu$m on
a cylinder of height $1750\mu$m. The emitter mounted on a cathode plate is grounded while
the anode plate is at a height $10000\mu$m above the cathode and kept at a positive potential $V$.
The parabolic cone has an apex radius of curvature $R_a$ in the range 5-30nm and
the cone half-angle is in the range $3^\circ-6^\circ$ as described in [\onlinecite{cabrera}].

The scaling behaviour can thus be studied using Eqns.~(\ref{eq:approxgam}) and (\ref{eq:tanh1})
with $D = h + d$. Note that since voltage scaling is being studied, the apex
field may be expressed as $\gamma_a(D)~V/D$ so that the appropriate quantity to study
is $\gamma_a(D)/D$.

\begin{figure}[hbt]
  \begin{center}
    \vskip -0.5cm
\hspace*{-1.0cm}\includegraphics[scale=0.35,angle=0]{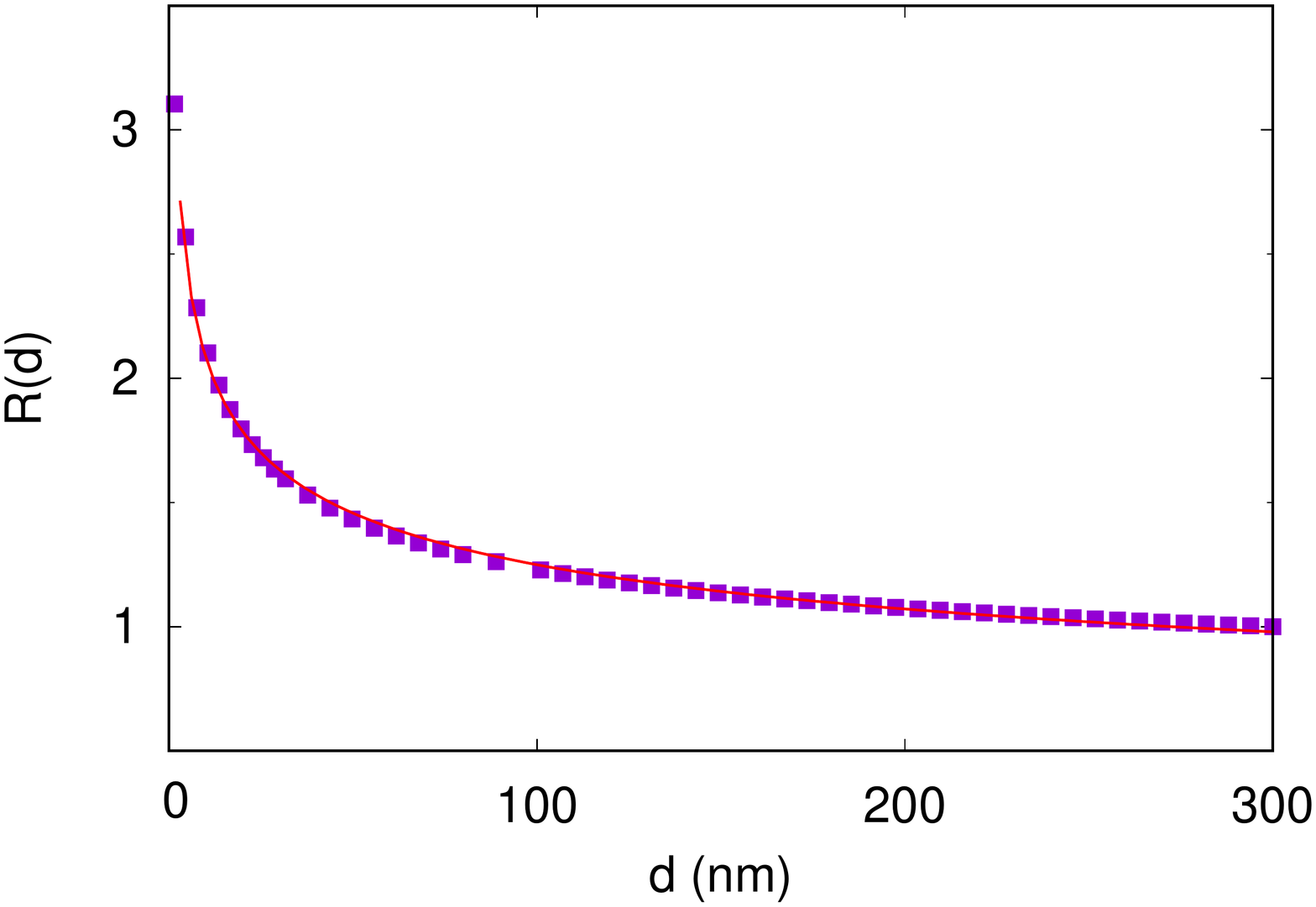}
\caption{The scaling function $R(d)$ of Eq.~(\ref{eq:Rd}) in the range 3-300nm for $R_a = 20$nm for
  cone half-angle $3.9^\circ$. Also shown is a power law fit $R(d) \sim d^{-\lambda}$ as a continuous curve 
with $\lambda \simeq 0.22$.}
\label{fig:Rd}
\end{center}
\end{figure}

In the range 3-300nm of $d$, the scaling behavior was studied by collapsing all other $I-V$ curves onto
the $d_{max}=300$nm  curve, by multiplying the voltage with a number $R(d)$. Thus, in order that the
apex field (hence the tunneling potential) be identical, the factor scaling the potential must be

\be
R(d) = \frac{\gamma_a(D)}{\gamma_a(D_{max})}\frac{D_{max}}{D}  \label{eq:Rd}
\ee

\noi
where $D_{max} = h + d_{max}$ and $D = h + d$.

Since the experimental situation has some uncertainties in the value of $R_a$ and cone-angle,
the simulation was carried out by varying these parameters such that the shape and values
of $R(d)$ are close to those reported in [\onlinecite{cabrera}] for the range of $d$ considered.
Fig.~\ref{fig:Rd} shows a plot of $R(d)$ as calculated using Eq.~(\ref{eq:Rd}) (denoted by solid squares) for $R_a = 20$nm
and cone half angle $3.9^\circ$.
Also shown is the best power law fit as a continuous curve. It is found that $R(d) \sim d^{-0.22}$
which is in agreement with the result reported in the experiment\cite{cabrera}.
Note that, the values of the cone-angle and apex radius chosen for Fig.~\ref{fig:Rd} are only
indicative since an eye-estimation
has been used in determining the closeness to $R(d)$ reported in [\onlinecite{cabrera}].
Importantly, the theory presented here allows
the existence of scaling behaviour well within the range of $R_a$ and cone-angle
reported in the experiment.

At very large $d$, $\gamma_a(D)$ saturates and $D \simeq d$ so that $R(d) \sim d^{-1}$. The large
$d$ behavior thus  depends on the range chosen with respect to the height of the emitter.

\section{Discussion and Conclusions}

The line charge model has proved to be a useful tool for vertical emitters aligned in the direction of
the asymptotic electric field. Though the initial study in this direction assumed the presence of an anode
at a finite distance $D$ for an ellipsoidal emitter\cite{pogo2009} (linear line charge), it has frequently been
used assuming a nonlinear line charge distribution \cite{jap16,db_fef,db_distrib,cosine,db_ultram}
and the anode to be far away so that its presence can
be neglected \cite{harris15,jap16,db_fef,db_distrib,db_rudra,db_ultram,cosine}.
In the preceding sections, we have investigated two of these results considering
the anode to be at a finite distance.

The first result concerns the apex field enhancement factor (AFEF). The introduction of the anode leads to
a increase in enhancement which can be significant when $D < 3h$ where $h$ is the height of the emitter.
The quantum of increase in AFEF depends on $h/R_a$ and is large for sharper emitters.
Importantly, it was numerically demonstrated using COMSOL that the anode proximity term can be treated
as a geometric effect and can be plugged into the expression for the anode-at-infinity AFEF.

The second result deals with the variation of the field enhancement factor on the emitter surface
but close to the apex. This is important from the point of view of field emission. Previously
derived results show that when the anode is far away, the field enhancement varies on the surface
following a generalized cosine law. Our results show that while the enhancement factor does
increase in the presence of the anode, it continues to obey the generalized cosine law.
This was again confirmed numerically using COMSOL.
These two results also explain an experimentally observed scaling behaviour of the $I-V$ curve accurately
when the anode is in close proximity to the emitter tip.

While the analysis is valid for a wide range of emitter shapes as demonstrated numerically, there are
exceptions. The elliposoid on cylindrical post (ECP) has the hemisphere on cylindrical post (HCP) as
a limiting case when the ellipsoid height $h_e$ equals the apex radius $R_a$. Modeling of ECP shapes where
$h_e < 5R_a$ does not yield results consistent with the analysis presented here. It is also
well known that an HCP does not obey the
generalized cosine law with the surface electric field falling slowly away from the apex. While, the
existence of a nonlinear line charge holds, the expansion of various quantities using partial integration
is fraught with difficulties as $f'(z)$ becomes large in the region where the cylinder makes a transition to an ellipsoid
for cases where $h_e$ approaches $R_a$. Apart from the HCP, other emitter shapes
may be thought of where the behaviour of
the line charge density changes abruptly. Despite these exceptions, the results presented here hold
for generic smooth emitter shapes and have relevance for experiments.

\vskip 0.25 in
\section{Acknowledgement}

The author thanks Rajasree Ramachandran, Gaurav Singh, Raghwendra Kumar and Rashbihari Rudra for
useful discussions. The author also acknowledges help from
Raghwendra Kumar and Shreya G. Sarkar in the use of COMSOL Multiphysics.

\vskip -0.25 in
$\;$\\
\section{References} 


\end{document}